\documentclass[conference,10pt]{IEEEtran}

\usepackage[ruled,vlined]{algorithm2e}
\usepackage{float}
\usepackage{graphicx}
\usepackage{subcaption}
\usepackage[T1]{fontenc}
\usepackage[table,xcdraw]{xcolor}
\usepackage{multirow}
\usepackage{dblfloatfix}
\usepackage{caption}
\usepackage[numbers,sort&compress]{natbib}
\usepackage{url}
\usepackage{enumitem}
\usepackage{amsmath}
\usepackage{amssymb}
\usepackage{breqn}
\usepackage{footnote}
\usepackage[symbol]{footmisc}

\SetAlgorithmName{Procedure}{procedure}{List of Procedures}

\begin{document}

\title{SACRIFICE: A Secure Road Condition Monitoring Scheme over Fog-based VANETs}

\author{\IEEEauthorblockN{Nishttha Sharma}
\IEEEauthorblockA{Indian Institute of Engineering \\ Science and Technology\\
Howrah, India\\
Email: nishttha.sharma@gmail.com}
\and
\IEEEauthorblockN{Jayasree Sengupta}
\IEEEauthorblockA{Indian Institute of Engineering \\ Science and Technology\\
Howrah, India\\
Email: jayasree202@gmail.com}
\and
\IEEEauthorblockN{Sipra Das Bit}
\IEEEauthorblockA{Indian Institute of Engineering \\ Science and Technology \\ Howrah, India \\ Email : sdasbit@yahoo.co.in}}

\maketitle

\begin{abstract}
    With the rapid growth of Vehicular Ad-Hoc Networks (VANETs), huge amounts of road condition data are constantly being generated and sent to the cloud for processing. However, this introduces a significant load on the network bandwidth causing delay in the network and for a time-critical application like VANET such delay may have severe impact on real-time traffic management. This delay maybe reduced by offloading some computational tasks to devices as close as possible to the vehicles. Further, security and privacy of vehicles is another important concern in such applications. Thus, this paper proposes a secure road condition monitoring scheme for fog-based vehicular networks. It considers Road-Side Units (RSUs) with computational capabilities, which act as intermediate fog nodes in between vehicles and cloud to reduce delay in decision making and thereby increases scalability. Apart from fulfilling basic security features, our scheme also supports advanced features like unlinkability and untraceability. A detailed security analysis proves that the proposed scheme can handle both internal and external adversaries. We also justify the efficiency of our scheme through theoretical overhead analysis. Finally, the scheme is simulated using an integrated SUMO and NS-3 platform to show its feasibility for practical implementations while not compromising on the network performance. The results show an average 80.96\% and 37.5\% improvement in execution time and end-to-end delay respectively while maintaining at par results for packet delivery ratio over a state-of-the-art scheme.
\end{abstract}
\IEEEpeerreviewmaketitle

\section{Introduction}

With the growth of the modern society along with the rise of smart cities, the number of connected devices and vehicles have increased exponentially. This has prompted the development of Vehicular Ad-Hoc Networks (VANETs) which not only facilitates traffic management but also makes driving safer by exploiting the inter-vehicular communication feature \cite{Lin_08}. As a result of this, huge amounts of data are being generated every second which needs to be processed in a time-sensitive manner. This has inspired the introduction of fog computing \cite{SRD_TII'21} in VANETs to reduce transmission delay, network bandwidth and improve overall traffic management. By extending the cloud closer to the vehicles in the network, fog has been able to reduce delay in decision making thereby improving real-time road condition monitoring, real-time driving assistance etc \cite{CHW_IoT'19}. However, introduction of fog in VANETs may incur additional security threats apart from the inherent ones, including misuse or leakage of sensitive data which needs to be handled with utmost care.

The existing literature have focused on introducing fog as a middleware in VANETs to utilise its benefits. The works \cite{CSM'19, Dong_IoT'20} have introduced a layer of fog devices between vehicles and cloud to provide improved network throughput, reduced latency and increased scalability. However, even after introducing fog several security and privacy issues, data quality and scheduling related challenges still exists. To address these concerns, the existing works focus on managing various security aspects in VANETs. The works \cite{BLS_IoT'17,MEW_IoT'19,ASS-TVT'18,CHW_IoT'19} propose different methods for mutual authentication by using securely agreed session keys and/or certificateless aggregate signcryption techniques to support privacy protection. But, it has high computation-communication overheads. On the contrary, the works \cite{Access'18,LLO_Systems'20} have proposed an identity-based mutual authentication protocol with only hash functions and Ex-OR operations which reduces overheads significantly. Recently, the works \cite{Islam_FGCS'18,CUI_VC'18} have proposed an authentication protocol for VANETs with password and group key agreement. However, most of these works only concentrate on securing the vehicle to RSU communication. Finally, the work \cite{WDU_TIFS'19} has proposed a secure road condition monitoring scheme with authorized reporting, privacy-preserving monitoring, and source authentication. However, their scheme is cloud-based which causes increased latency for a system like VANET. It also does not take into consideration the privacy of the vehicle user and a few other security features like resistance to various attacks such as man-in-the-middle, replay etc. 

From the above discussion it is clear that the state-of-the-art works rely largely on a cloud-based platform which results in a higher end-to-end delay. Further, despite the fact that these works focus on certain security features like user anonymity, mutual authentication, non-repudiation etc. various other important security features still remain less investigated in such VANET environments. This motivates us to design an improved architecture for VANETs and propose a \textbf{S}ecure ro\textbf{A}d \textbf{C}ondition monito\textbf{RI}ng scheme over \textbf{F}og-based veh\textbf{IC}ular n\textbf{E}tworks (\textbf{SACRIFICE}) keeping in mind the application's time-sensitive nature. The SACRIFICE also targets to incorporate security features (e.g. unlinkability to prevent attackers from tracking the movement of vehicles) in addition to basic security while running the scheme. Therefore, the contributions put forth by our work are as follows:

\begin{itemize}[leftmargin=*]
    \item We propose SACRIFICE having the following features:
    \begin{itemize}
        \item reduce delay in decision making by considering Fog-based VANET while maintaining basic security features like mutual authentication, user anonymity etc.
        \item introduce additional security features like non-repudiation, unlinkability and untraceability.
    \end{itemize}
    \item A detailed security analysis proves that SACRIFICE can handle both internal and external adversaries.
    \item We validate SACRIFICE both theoretically and experimentally.
    \begin{itemize}
        \item Establish SACRIFICE to be lightweight as well as having low overheads compared to state-of-the-art works.
        \item Simulation results in an integrated real-time platform using SUMO and NS-3 establish the practicality of the scheme.
    \end{itemize}
\end{itemize}

The rest of the paper is structured as follows. Section II discusses the system model. SACRIFICE is presented in Section III. Section IV briefly explains the security analysis of the scheme. Section V highlights the performance of the scheme. Finally, Section VI concludes the work.

\section{System Model}

This section illustrates the system model in detail where the fog-based VANET architecture inspired from \cite{WDU_TIFS'19,CHW_IoT'19} is used as the backbone of our work. Followed by that, we explain the security guarantees and the adversarial model.

\subsection{Architecture}

The architecture shown in Fig. \ref{fig:Image1} comprises of four layers and has inculcated the advantageous features of \cite{WDU_TIFS'19,CHW_IoT'19}. Each of these layers has entities like vehicles, RSUs etc. and can communicate with its immediate upper and lower layers. The activities of these four layers are described below:

\begin{figure}[!ht]
\centering
\fbox{\includegraphics[scale=0.10]{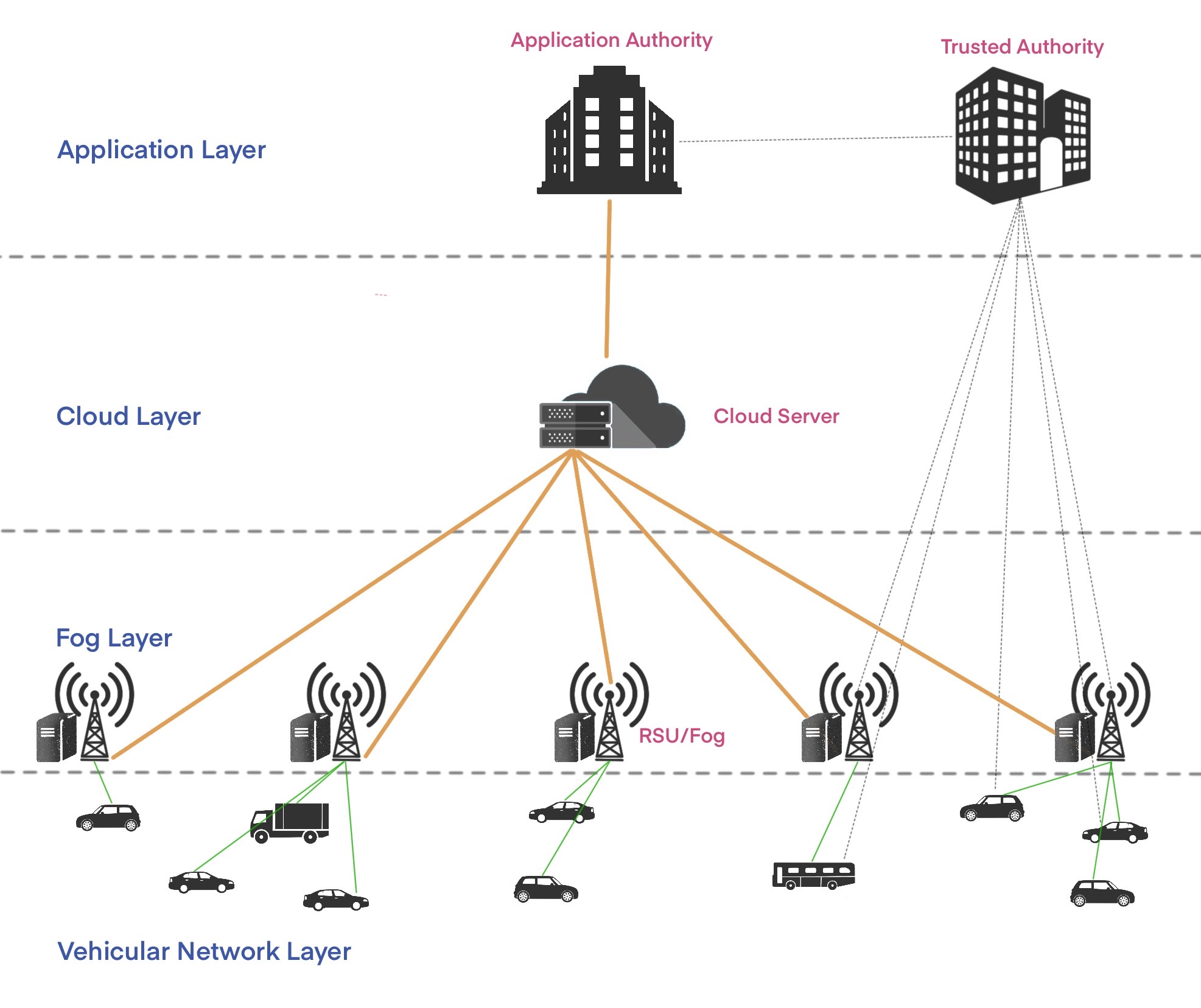}}
\caption{\small \sl Fog-based VANET Architecture}
\label{fig:Image1}
\end{figure}

\noindent \textbf{Vehicular Network Layer:} It consists of vehicles equipped with various sensors (camera, temperature, etc.) \cite{BDB_WN'19} and On-Board Units (OBUs) which has communication capabilities. The vehicles are responsible for gathering the sensory data and sending this information along with the location and time to the fog layer. In turn the vehicles receive information from fog nodes in case of a service request.

\noindent \textbf{Fog Layer:} It consists of Road-Side Units (RSUs) enabled with computing capabilities. These are installed at important junctions of the roads maintaining specific distance from each other depending upon the communication range to provide maximum coverage while guaranteeing persistent links with the cloud. RSUs use Dedicated Short-Range Communication to interact with the vehicles. These nodes are more robust and are responsible for minimizing the delay in decision making by extending the cloud closer to the vehicular network layer \cite{SRD_TII'21}. An RSU receives sensory data from various vehicles within its range, processes it and sends it to the cloud for further action.

\noindent \textbf{Cloud Layer:} It typically consists of different kind of storage or application servers which are responsible for communicating with the fog layer in order to receive data about the entire network. The cloud may process such data and forward certain reports to the Application Authority, if required.

\noindent \textbf{Application Layer:} It consists of the Trusted and the Application Authorities. The Trusted Authority is responsible for device registration (e.g. vehicles) and key distribution whereas Application Authority takes necessary actions based on processed data. For example, it can ask the vehicles to take a different route  in case of congestion or accidents.  

\subsection{Security Guarantees and Adversarial Model}

The security features, adversarial model and assumptions considered in SACRIFICE are discussed here.

\noindent \textbf{Security of the Scheme}

The following are the security requirements of the scheme which are taken care of:

\noindent \textbf{Mutual Authentication \cite{MEW_IoT'19,CHW_IoT'19}:} The validity of all participants (i.e. vehicles, RSUs) needs to be guaranteed. This requires that none of the malicious participants should be able to impersonate some other valid participant without being detected. Thus, vehicles and RSUs should authenticate each other to prevent forgery of tokens exchanged between them.
    
\noindent \textbf{User Anonymity and Untraceability \cite{MEW_IoT'19,CHW_IoT'19}:} It requires protecting the vehicle users' privacy during data transmission to hide its real identity and behavioural patterns from the network. This means that an attacker intercepting the messages cannot extract a user's real identity or track its behaviour (e.g. route, driving patterns).
    
\noindent \textbf{Non-Repudiation \cite{CHW_IoT'19}:} The scheme guarantees that vehicles shouldn't be able to deny its involvement in case of any dispute (e.g. sending corrupted data). It implies that in case of denial from the vehicles, the RSUs will be able to prove the role of the vehicles to any third party.
    
\noindent \textbf{Unlinkability \cite{LLO_Systems'20}:} This feature prevents an adversary from determining whether two messages ($m_1$, $m_2$) have originated from the same vehicle or not. Thus, an adversary will be unable to link messages generated by the same vehicle and thereby fail to distinguish between vehicles.
    
\noindent \textbf{Resistance to common attacks \cite{MEW_IoT'19}:} To ensure security of the scheme, it is important to prevent attacks like man-in-the-middle and replay attacks \cite{Ghosal_ICCSA'10}. For example, an attacker neither can pretend to be a legitimate user to cheat other participants nor can it launch an attack on the scheme to continue sending old messages to overload the network. 

\noindent \textbf{Adversarial Model}

\noindent \textbf{Entity:} An entity can either be honest, semi-honest or malicious. Semi-honest entities do not deviate from the protocol specifications but may intend to obtain intermediate results/information from the nearby entities. On the contrary, malicious entities may deviate from the protocol arbitrarily.

\noindent \textbf{Adversary:} An adversary is a polynomial-time algorithm that can compromise any party at any point of time, subject to some upper bound \cite{arxiv_BNR'19}. Adversaries can be broadly categorized into two types: internal and external. \textit{External adversaries} do not possess authentic keying material and hence they cannot participate as valid nodes \cite{SRD_FICN'18}. They can only eavesdrop on radio transmissions and try to access information from the data transmitted through the channels. On the contrary, \textit{internal adversaries} possess authentic keying material and have more effective and powerful resources in terms of energy and communication capabilities and are more vulnerable than external adversaries \cite{SRD_FICN'18}. When an internal adversary captures a device in the network, it means the adversary has gained control over the tokens (not stored in a tamper-proof box) in the device. It also gains control over the messages sent/received by the device. We assume an adversary can neither interfere with the message exchanges between honest parties nor can it break cryptographic primitives like hash functions, except with a negligible probability. In this work, we consider both external and internal adversaries, however they have bounded computational and storage capabilities.

\noindent \textbf{Assumptions:} The following assumptions are made while setting up the proposed scheme:

\begin{itemize}[leftmargin=*]
	\item There is a secure channel between the Application Authority and the Trusted Authority as well as between the Trusted Authority and a device that is being registered.
	\item The vehicles can be malicious, RSUs are semi honest and Cloud is an untrusted entity.
	\item The Application Authority and the Trusted Authority are honest and trusted entities.
\end{itemize}

\section{Proposed Scheme}

A detailed overview of our proposed \textbf{S}ecure ro\textbf{A}d \textbf{C}ondition monito\textbf{RI}ng scheme over \textbf{F}og-based veh\textbf{IC}ular n\textbf{E}tworks (\textbf{SACRIFICE}) along with its algorithmic constructs is discussed in this section. The scheme generates a road condition report through a distributed process running in the participating vehicles and the roadside units of the network with the RSUs performing the intensive computations. 

\noindent \textbf{Working Principle:} Whenever a vehicle $U_i$ enters the scope of an RSU $R_j$, both the devices have to mutually authenticate each other. After successful authentication, $U_i$ sends an initial road condition report to $R_j$ from which it generates the final road condition report. This final report is sent to the cloud for storage and further processing. The report can be extracted by the Application Authority to make important decisions in case of an emergency. To ensure honest behaviour of the participants, a hash-based lightweight mutual authentication algorithm has been implemented. Our proposed scheme consists of five different phases which are outlined below. 

\begin{figure}[!ht]
\begin{center}
\fbox{\footnotesize
\begin{minipage}{0.95\columnwidth}
\begin{center}
    \underline{\textbf{SACRIFICE}}\\
\end{center}
\begin{itemize}[leftmargin=*]
    \item \textbf{System Setup:}
    The Application Authority (AA) chooses a $q-$order additive group $\mathbb{G}$ with generator $P$ and does the following:
    \begin{itemize}
        \item Generates secret key $msk \in \mathbb{Z}_q^*$, public key $P_{pub}=msk \cdot P$ and alert threshold $\tau$.
        \item Sends $msk$ to the Trusted Authority (TA).
        \item Chooses cryptographic hash functions $h_1,h_2,h_3,h_4,h_5,h_6,h_7$.
        \item Publishes the public parameters:\\
$prms \: = \: (q,\mathbb{G},P,P_{pub},h_1,h_2,h_3,h_4,h_5,h_6,h_7,\tau)$.
    \end{itemize}
        \item \textbf{Device Registration:} The vehicles and the RSUs registers themselves with the system via a secure channel.
        \begin{itemize}[leftmargin=*]
            \item $U_i$ and $R_j$ sends their identities $ID_{U_i}$ and $ID_{R_j}$ to TA.
            \item TA generates keys $P_{U_i} \leftarrow \mathbf{genKey(ID_{U_i})}$ and $P_{R_j} \leftarrow \mathbf{genKey(ID_{R_j})}$   \tcp{Procedure 1}
            \item TA sends $P_{U_i}$ and $(P_{R_j},msk)$ to $U_i$ and $R_j$ respectively.
            \item $U_i$ stores $P_{U_i}$ and $R_j$ stores $(P_{R_j},msk)$.
        \end{itemize}
        \item \textbf{Mutual Authentication:} When a vehicle $U_i$ enters the scope of a RSU $R_j$, the mutual authentication step executes as below:
        \begin{itemize}
            \item $U_i$ computes $M_1\ \leftarrow\ \mathbf{authreq(ID_{U_i},P_{U_i},ID_{R_j})}$. \tcp{Procedure 2}
            \item Sends $M_1$ to $R_j$ for authentication. 
            \item $R_j$ on receiving $M_1$ from $U_i$, executes $isValidReq\ \leftarrow\ \mathbf{authres(M_1,msk,ID_{R_j})}$ \tcp{Procedure 3}
            \item $U_i$ calls $isValidRes \leftarrow \mathbf{authack (M_2,ID_{U_i},P_{U_i},ID_{R_j})}$ on receiving $isValidReq$ as $true$ \tcp{Procedure 2}
            \item $U_i$ sets $mutualAuth = true$, on receiving $isValidRes$ as $true$.
             \end{itemize}
        \item \textbf{Report Generation:} When vehicle $U_i$ gathers some road condition information $I$, it does the following:
        \begin{itemize}
            \item Generates report $M_3 \leftarrow \mathbf{initialReport(I,snky,ID_{U_i},}$ $\mathbf{P_{U_i},ID_{R_j})}$. \tcp{Procedure 2}
            \item Sends $M_3$ to $R_j$.
            \item On receiving $M_3$, $R_j$ executes $M_4 \leftarrow \mathbf{finalReport (M_3,snky,ID_{R_j},P_{R_j},msk)}$ \tcp{Procedure 3}
        \end{itemize}
    \item \textbf{Report Processing:} When Cloud Server (CS) receives final report $M_4$, it does the following:
        \begin{itemize}[leftmargin=*]
            \item Generates $reportAA \leftarrow \textbf{processCS (\{L,W\})}$ \tcp{Procedure 4}
            \item For a correct report, CS sends $reportAA$ to Application Authority (AA).
            \item AA on receiving $reportAA$ from CS calls $reportAccepted \leftarrow \textbf{processAA (reportAA)}$ \tcp{Procedure 5}
            \item $AA$ accepts $reportAA$ as valid and takes necessary actions when $reportAccepted$ is $true$.
        \end{itemize}
    \end{itemize}
\end{minipage}
}
\setlength{\belowcaptionskip}{-10pt}
\caption{\small \sl Detailed description of SACRIFICE for honest participants \label{fig:Image2}}
\end{center}  
\end{figure}

\noindent \textbf{Phase 1: System Setup}

This phase is performed during the establishment of the network. In this phase, each of the parties involved in the communication converge on the security parameters $prms \: = \:(q,G,P,P_{pub},h_1,h_2,h_3,h_4,h_5,h_6,h_7,\tau)$.

\noindent \textbf{Phase 2: Device Registration} 

All the devices, i.e. both vehicles and RSUs are registered prior to entering the network.

\noindent \textbf{Phase 3: Mutual Authentication} 

The mutual authentication between the vehicles and the RSUs is inspired from \cite{MEW_IoT'19,LLO_Systems'20}, however we have modified their schemes to eliminate interactions with the Trusted Authority (TA) during this phase. This reduces the trust dependence on any third party entity during the execution of SACRIFICE.

When a vehicle $U_i$ enters the scope of an RSU $R_j$, the $U_i$ sends out the required tokens as authentication request to the $R_j$. On receiving the authentication request, RSU checks its validity. For a successful validation, $R_j$ generates the required tokens as authentication response and sends it to $U_i$. In turn, $U_i$ checks the validity of the authentication response. On successful validation it is said that both the vehicle and the RSU have mutually authenticated each other and can proceed further. On failure, of any of the above steps, the protocol is terminated. 

\begin{algorithm}[!htb]
\small
\DontPrintSemicolon
\SetKwProg{Fn}{Function}{:}{end}
\footnotesize
\Fn{genKey (ID)}{
Calculate $P \: = \: h_1 (msk,ID)$\\
\KwRet $P$
}
\caption{Executed by Trusted Authority (TA)}
\label{algo:Algo1}
\end{algorithm}

\begin{algorithm}[!htb]
\small
\DontPrintSemicolon

\SetKwProg{Fn}{Function}{:}{end}
\footnotesize
\Fn{authreq ($ID_{U_i}$, $P_{U_i}$, $ID_{R_j}$)}{
Generate a random variable $r_i \in Z_q^*$\\
Calculate and store the following tokens:\\
$X_1 \: = \: r_i \cdot P$\\
$X_2 \: = \: r_i \cdot P_{pub}$\\
$X_3 \: = \: h_2 (X_1,X_2,t_{U_i}) \oplus ID_{U_i}$ and\\
$C_i \: = \: h_3 (ID_{U_i} ID_{R_j},P_{U_i},X_1,X_2,t_{U_i})$\\
\KwRet $M_1 \: = \: \{X_1,X_3,C_i,t_{U_i}\}$
}

\Fn{authack ($M_2$, $ID_{U_i}$, $P_{U_i}$, $ID_{R_j}$)}{

\If{($t'_{R_j} \: - \: t_{R_j} \leq \triangle t$)}{
	\If{($C_j \: = \: h_4 (ID_{U_i},P_{U_i},ID_{R_j},Y_1,Y_1 \cdot X_2,t_{R_j})$)}{
		Calculate $snky \: = \: r_i \cdot Y_1$\\
		\KwRet $true$
	}
}
}

\Fn{initialReport ($I$, $snky$, $ID_{U_i}$, $P_{U_i}$, $ID_{R_j}$)}{
Calculate tokens $Q_1 \: = \: h_5 (\widehat{t}_{U_i}, snky, ID_{R_j}) \oplus ID_{U_i}$ and $Q_2 \: = \: ID_{R_j} \oplus I \oplus P_{U_i}$\\
\KwRet ($M_3 \: = \: \{Q_1,Q_2,\widehat{t}_{U_i}\}$)

}

\caption{Executed by Vehicle $U_i$}
\label{algo:Algo2}
\end{algorithm}

\begin{algorithm}[!htb]
\small
\DontPrintSemicolon

\SetKwProg{Fn}{Function}{:}{end}
\footnotesize
\Fn{authres ($M_1$, $msk$, $ID_{R_j}$)}{
\If{($t'_{U_i} \: - \: t_{U_i} \leq \triangle t$)}{
	Calculate $ID_{U_i} \: = \: X_3 \oplus h_2 (X_1,X_1 \cdot msk,t_{U_i})$\\	
	\If{($C_i \: = \: h_3 (ID_{U_i}, ID_{R_j}, h_1 (msk, ID_{U_i}), X_1,$ $X_1\cdot msk, t_{U_i})$)}{
		Generate a random variable $r_j \in Z_q^*$\\
		Calculate the following tokens:\\ 
		$Y_1 \: = \: r_j \cdot P$\\
		$Y_2 \: = \: r_j \cdot X_1 \cdot P_{pub}$\\
		$C_j \: = \: h_4 (ID_{U_i}, h_1 (msk, ID_{U_i}), ID_{R_j}, Y_1,$ $Y_2, t_{R_j})$ and
		$snky \: = \: r_j \cdot X_1$.\\
		Store $snky$ and send $M_2 \: = \: \{Y_1,C_j,t_{R_j}\}$ to $U_i$
		\KwRet $true$
	}
}
\KwRet $false$.
}

\Fn{finalReport ($M_3$, $snky$, $ID_{R_j}$, $P_{R_j}$, $msk$)}{
\If{($\widehat{t}'_{U_i} \: - \: \widehat{t}_{U_i} \leq \triangle t$)}{
	Generate a random variable $s \in Z_q^*$.\\
	Calculate the tokens $ID_{U_i} \: = \: Q_1 \oplus h_5 (\widehat{t}_{U_i}, snky,$ $ID_{R_j} )$ and
	$I \: = \: Q_2 \oplus ID_{R_j} \oplus h_1 (msk, ID_{U_i})$\\
	Calculate the final report $\{L,W\}$ where $L \: = \: (l_1, l_2, l_3, l_4, l_5)$ and $W \: = \: (\widehat{t}_{U_i}, \widehat{t}_{R_j})$\\
	$l_1 \: = \: s\cdot P$\\
	$l_2 \: = \: s \cdot h_6 (ID_{R_j}, P_{R_j}, I)$\\
	$l_3 \: = \: h_6 (msk \cdot P) \oplus ID_{R_j}$\\
	$l_4 \: = \: h_6 (ID_{R_j}, msk) \oplus I$ and\\
	$l_5 \: = \: h_7 (l_1,l_2,l_3,l_4,\widehat{t}_{U_i},\widehat{t}_{R_j})$\\
	\KwRet $M_4 \: = \: \{L,W\}$
}
}

\caption{Executed by RSU $R_j$}
\label{algo:Algo3}
\end{algorithm}

\noindent \textbf{Phase 4: Report Generation}

We adopt report generation and processing method from \cite{WDU_TIFS'19}. But unlike \cite{WDU_TIFS'19}, here we divide the report generation task between the RSUs and vehicles instead of getting it performed by the vehicles alone. This report generation is broken down into two sub-tasks: \textit{initial report} and \textit{final report} generations respectively. This substantially reduces the delay of the entire scheme and improvises faster decision making by the Application Authority (AA).

In this phase a vehicle may generate an initial road condition report and send it to its nearest RSU. The RSU in turn processes the report to extract necessary information. It then generates a final road condition report which is sent to the Cloud Server (CS).

\begin{algorithm}[!htb]
\small
\DontPrintSemicolon

\SetKwProg{Fn}{Function}{:}{end}
\footnotesize
\Fn{processCS ($\{L,W\}$)}{
\If{($t'_{U_i} \: - \: t_{U_i} \leq \triangle t$ and $t'_{R_j} \: - \: t_{R_j} \leq \triangle t$)}{
	\eIf{($l_5 \: = \: h_7 (l_1,l_2,l_3,l_4,\widehat{t}_{U_i},\widehat{t}_{R_j})$)}{
		\For{(each equivalence class $G'$ in CS)}{
			Retrieve a report $\{L',W'\}$ from $G'$\\
			\If{($l_1 \cdot l'_2 \: = \: l'_1 \cdot l_2$)}{
				Insert $\{L,W\}$ in equivalence class $G'$\\
				\If{($|G| > \tau$)}{
					\KwRet $\{L,W\}$
				}
				\KwRet $false$
			}
		}
		\If{(no match was found)}{
			Create a new equivalence class to insert $\{L,W\}$.\\
			\KwRet $false$
		}
	}{
		Invalid Report Received from RSU\\
		Return $false$
	}
}
\KwRet $false$
}
\caption{Executed by Cloud Server (CS)}
\label{algo:Algo4}
\end{algorithm}

\begin{algorithm}[!htb]
\small
\DontPrintSemicolon

\SetKwProg{Fn}{Function}{:}{end}
\footnotesize
\Fn{processAA ($\{L,W\}$)}{

Calculate the following tokens:\\
$ID_{R_j} \: = \: l_3 \oplus h_6 (msk \cdot P)$ and\\
$I \: = \: l_4 \oplus h_6 (ID_{R_j},msk)$\\
\eIf{($l_1 \cdot h_6 (ID_{R_j},h_1 (msk,ID_{R_j}),I) \: = \: l_2 \cdot P$)}{
	\KwRet $true$
}{
	\KwRet $false$
}
}

\caption{Executed by Application Authority (AA)}
\label{algo:Algo5}
\end{algorithm}

\noindent \textbf{Phase 5: Report Processing}

When the Cloud Server receives a report from an RSU, it first checks the validity of the report and stores it in an appropriate equivalence class, if the report is valid. Here, equivalence class refers to a class consisting of a set of tuples which report the same road condition for the same location within a reasonable time period \cite{WDU_TIFS'19}. If the targeted equivalence class reaches a predefined threshold $\tau$, a report from that particular class is sent to the Application Authority (AA) for further processing. The AA in turn tests the validity of the report and based on it decides whether the report is to be accepted or not. If accepted, it extracts the road condition information from the report.

Fig. \ref{fig:Image2} gives a first level view of the entire scheme whereas the tasks performed by each of the entities mentioned in this figure are elaborated through the respective procedures provided next. Here, the functions executed by Trusted Authority (TA), Vehicle $U_i$, RSU $R_j$, Cloud Server (CS) and Application Authority (AA) are explained in detail in Procedures \ref{algo:Algo1}, \ref{algo:Algo2}, \ref{algo:Algo3}, \ref{algo:Algo4} and \ref{algo:Algo5} respectively. Referring to the figure and the procedures, at any instance of time many such $(U_i, R_j)$ pair can communicate. Hence, our scheme fits well to a multiple vehicle-RSU setting as well.

\section{Security Analysis}

This section analyzes the security of SACRIFICE.

\subsection{Mutual Authentication}

In SACRIFICE, when a malicious vehicle $U_i$ attempts to mutually authentication itself with a legitimate RSU $R_j$, the following computations are performed:

\noindent Let $U_i$ select a random $r_i \in Z_q^*$, a forged identity $ID_{U_i}$, and derive the current timestamp $t_{U_i}$. It then calculates the following:\\
	\hphantom{~~~~} $X_1 = r_i \cdot P$\\
	\hphantom{~~~~} $X_2 = r_i \cdot P_{pub}$\\
	\hphantom{~~~~} $X_3 = h_2 (X_1,X_2,t_{U_i}) \oplus ID_{U_i}$\\
	However, to calculate $C_i$ it must have the value of $P_{U_i}$, which is only sent to vehicles during registration. Without the value of $msk$, $U_i$ cannot calculate $P_{U_i}$. Let us assume that $U_i$ selects a random value $P'_{U_i}$ and computes,\\
	\hphantom{~~~~} $C_i = h_3 (ID_{U_i},ID_{R_j},P'_{U_i},X_1,X_2,t_{U_i})$\\
	and sends $M_1 = \{X_1,X_3,C_i,t_{U_i}\}$ as the authentication request to $R_j$. On receiving $M_1$, $R_j$ calculates:\\
	\hphantom{~~~~} $ID_{U_i} = X_3 \oplus h_2 (X_1,X_1 \cdot msk,t_{U_i})$
	and\\
	\hphantom{~~~~} $C_i' = h_3 (ID_{U_i},ID_{R_j},h_1 (msk,ID_{U_i}),X_1,X_1 \cdot msk,t_{U_i})$\\
	However, on checking $R_j$ finds that $C_i'$ is not equal to the received $C_i$ in $M_1$ (since $P'_{U_i} \neq h_1 (msk,ID_{U_i}))$. Thus, $R_j$ rejects the authentication request. Hence, it is proved that SACRIFICE doesn't allow any malicious vehicle to operate in the network.
	
\subsection{User Anonymity and Untraceability}

\subsubsection{User Anonymity}
An adversary $\mathcal{A}$ (internal or external) may try to obtain the real identity of vehicle $U_i$, where the following two cases may arise:

\noindent \textbf{\textit{Case 1:}} Let us assume, an external adversary $\mathcal{A}$ intercepts message $M_1$ and obtains the value of $t_{U_i}$. It also obtains the value of $X_3=h_2 (X_1,X_2,t_{U_i}) \oplus ID_{U_i}$ where $X_1=r_i \cdot P$ and $X_2=r_i \cdot P_{pub}$. To extract the real identity $ID_{U_i}$ of $U_i$ from $X_3$, $\mathcal{A}$ needs the value of $X_2$ which can be calculated either using $r_i$ or $msk$, both of which are unknown to $\mathcal{A}$ and also cannot be evaluated by it in polynomial time. Similarly, $\mathcal{A}$ also cannot extract $ID_{U_i}$ from $C_i=h_3 (ID_{U_i}, ID_{R_j}, P_{U_i}, X_1, X_2, t_{U_i})$ since it requires $r_i$ for the same, which is unknown to it.

\noindent\textbf{\textit{Case 2:}} Let us now consider that $\mathcal{A}$ is an internal adversary and it takes control over the RSU $R_j$. It is already known that $\mathcal{A}$ cannot acquire the keys $msk$ and $P_{R_j}$. It can however gain control over the messages sent/received by $R_j$. Thus, when $U_i$ sends the authentication request message $M_1$, $\mathcal{A}$ intercepts $M_1$ and computes $ID_{U_i}$ as below:\\
	\hphantom{~~~~} $ID_{U_i}=X_3 \oplus h_2 (X_1,X_1 \cdot msk,t_{U_i})$\\
	Since, $msk$ cannot be accessed by $\mathcal{A}$, the value of $ID_{U_i}$ cannot be calculated as well. Thus, $\mathcal{A}$ assumes a random value $msk'$ in place of $msk$ and calculates\\
	\hphantom{~~~~} $ID'_{U_i}=X_3 \oplus h_2 (X_1,X_1 \cdot msk',t_{U_i})$\\
	Since $\mathcal{A}$ controls $R_j$, therefore $R_j$ sends\\
	\hphantom{~~~~} $M_2=\{Y_1,C_j',ID_{R_j},t_{R_j}\}$ to $U_i$ where,\\
	\hphantom{~~~~} $C_j' = h_4 (ID'_{U_i},h_1 (msk',ID'_{U_i}),ID_{R_j},Y_1,Y_2,t_{R_j})$\\
	On receiving $M_2$ from $R_j$, $U_i$ calculates\\
	\hphantom{~~~~} $C_j=h_4 (ID_{U_i},P_{U_i},ID_{R_j},Y_1,Y_1 \cdot X_2,t_{R_j})$\\
	However the calculated $C_j$ value mismatches the $C_j'$ extracted from received $M_2$ (since $ID'_{U_i} \neq ID_{U_i}$ and $P_{U_i} \neq h_1 (msk',ID'_{U_i})$). Therefore, $U_i$ terminates the authentication process thereby preserving user anonymity.
	
\subsubsection{Untraceability} 

An adversary $\mathcal{A}$ attempting to know the behaviour of a participant $R_j$ is discussed here. The messages sent from $R_j$ are $M_2$ and $M_4$. Here, $M_2$ is dependent on the random values $r_i$ and $r_j$ as well as timestamps $t_{U_i}$ and $t_{R_j}$. $M_4$ is dependent on the session key $snky$ and also on timestamps $\widehat{t}_{U_i}$ and $\widehat{t}_{R_j}$. The value of these random variables and timestamps are different for each instance of the respective messages. Therefore, every time a message is sent, its contents change due to their dependency on these dynamic variables making it difficult to understand the behaviour of $R_j$.\\
For example, two vehicles entering the range of $R_j$ at different timestamps attempt mutual authentication with it. $R_j$ then chooses two random $r_j$ values, for authentication with vehicles $U_1$ and $U_2$. Thus, the contents of the message (say $M_2$) will be different for $U_1$ and $U_2$ not only because they are sent out at different timestamps but also because of the presence of different random values in them.
	
\subsection{Non-Repudiation}

During mutual authentication, $R_j$ can calculate the real identity of vehicle $U_i$ from message $M_1$ as:\\
	\hphantom{~~~~} $ID_{U_i}=X_3 \oplus h_2 (X_1,X_1 \cdot msk,t_(U_i ))$;\\
	and also, from $M_3$ as:\\
	\hphantom{~~~~} $ID_{U_i}=Q_1 \oplus h_5 (\widehat{t}_{U_i},snky,ID_{R_j})$\\
	This is useful to prove the involvement of a vehicle, if the vehicle denies participation in case of any malicious behavior.
	
\subsection{Unlinkability}

When a vehicle $U_i$ first enters the range of an RSU $R_j$ and after sometime, in the range of another RSU $R_k$, an authentication request is sent by $U_i$ to the respective RSUs. $U_i$ sends $M_1$ and $M'_1$ to $R_j$ and $R_k$ respectively. We consider both these authentication requests have been fulfilled successfully. According to SACRIFICE, $U_i$ will choose a new random value for $r_i$ before sending out a new authentication request. Therefore, the tokens in $M_1$ dependent on these values will also change with every authentication request.\\
For example, if $M_1=\{X_1,X_3,C_x,t_{U_i}\}$ then $M_1' = \{X'_1, X'_3,C'_x,t'_{U_i}\}$, i.e., all the tokens of the authentication request message will change. So, when an attacker $\mathcal{A}$ intercepting both the messages $M_1$ and $M'_1$ from a public channel, cannot link the messages to the same vehicle as the contents of both the messages are different.

\subsection{Resistance to common attacks}

In this section, we discuss resistance against two of the most common attacks: (a) Man-in-the-Middle (b) Replay Attacks.

\subsubsection{Man-in-the-Middle Attacks} Here, we assume that an adversary $\mathcal{A}$ (internal or external) already knows the values of $ID_{U_i}$ and $ID_{R_j}$ and wishes to forge a message. The following two cases may arise:

\noindent\textbf{\textit{Case 1:}} When an external adversary $\mathcal{A}$ tries to forge message $M_2$, it chooses some random value $r'_j \in Z_q^*$ and calculates $Y_1=r'_j \cdot P$ and $Y_2=r'_j \cdot X_1 \cdot P_{pub}$. However, in order to calculate $C_j$ it requires the value of the master key $msk$ which is unknown to $\mathcal{A}$. Moreover, $C_j$ cannot be calculated in polynomial time. Hence it assumes a random value $msk'$ in place of $msk$ and computes $C_j'$ as below:\\
	\hphantom{~~~~} $C_j'=h_4 (ID_{U_i},h_1 (msk',ID_{U_i}),ID_{R_j},Y_1,Y_2,t_{R_j})$\\
	On receiving $M_2$, $U_i$ computes $C_j$ and finds out that the calculated $C_j$ is not equal to $C_j'$ received in $M_2$ (since $P_{U_i} \neq h_1 (msk',ID'_{U_i})$) and thereby terminates the authentication process. Thus, it is proved that an external adversary $\mathcal{A}$ cannot successfully forge message $M_2$.
	
\noindent\textbf{\textit{Case 2:}} We consider an internal adversary $\mathcal{A}$ has gained control over RSU $R_j$ and tries to forge messgae $M_2$. As per SACRIFICE, we already know that $\mathcal{A}$ cannot acquire the keys $msk$ and $P_{R_j}$ but can control all the messages sent/received by $R_j$. $\mathcal{A}$ chooses some random value $r'_j \in Z_q^*$ and calculates $Y_1=r'_j \cdot P$ and $Y_2=r'_j \cdot X_1 \cdot P_{pub}$. It then assumes a random $msk'$ in place of $msk$ and computes $C_j'$ as follows:\\
	\hphantom{~~~~} $C_j'=h_4 (ID_{U_i},h_1 (msk',ID_{U_i}),ID_{R_j},Y_1,Y_2,t_{R_j})$\\
	Since $\mathcal{A}$ controls the messages being sent out from $R_j$, therefore $R_j$ sends $M_2=\{Y_1,C_j',ID_{R_j},t_{R_j}\}$ to $U_i$. $U_i$ on receiving $M_2$ calculates $C_j$ as below:\\
	\hphantom{~~~~} $C_j=h_4 (ID_{U_i},P_{U_i},ID_{R_j},Y_1,Y_1 \cdot X_2,t_{R_j})$\\
	$U_i$ finds that the calculated $C_j$ is not equal to $C_j'$ received in $M_2$ (since $P_{U_i} \neq h_1 (msk',ID'_{U_i})$) and terminates the authentication process. Therefore, $\mathcal{A}$ cannot successfully forge message $M_2$ in this case as well.

\subsubsection{Replay Attacks} When an attacker $\mathcal{A}$ repeats or delays a message (say $M_3$), it reaches the recipient (i.e., RSU $R_j$) at timestamp $\widehat{t}'_{U_i}$. After receiving the message, $R_j$ will first verify the validity of the message by checking the freshness of timestamp $\widehat{t}_{U_i}$, i.e., whether\\
\hphantom{~~~~} $\widehat{t}'_{U_i} - \widehat{t}_{U_i} \leq \triangle t$\\
if not, then the session is terminated. This makes the protocol resistant to replay attack.

\begin{table}[!ht]
\centering
\caption{A Comparative Summary of Key Features}
\label{tab:Table5}
\scalebox{0.75}{%
\begin{tabular}{|c|c|c|c|c|c|}
\hline
\textbf{Features} & \textbf{\cite{WDU_TIFS'19}} & \textbf{\cite{MEW_IoT'19}} & \textbf{\cite{CHW_IoT'19}} & \textbf{\cite{LLO_Systems'20}} & \textbf{SACRIFICE} \\ \hline
Mutual Authentication & $\checkmark$ & $\checkmark$ & $\checkmark$ & $\checkmark$ & $\checkmark$ \\ \hline
User Anonymity & $\checkmark$ & $\checkmark$ & $\checkmark$ & $\times$ & $\checkmark$ \\ \hline
Untraceability & $\checkmark$ & $\times$ & $\checkmark$ & $\times$ & $\checkmark$ \\ \hline
Non-Repudiation & $\checkmark$ & $\checkmark$ & $\times$ & $\times$ & $\checkmark$ \\ \hline
Unlinkability & $\times$ & $\times$ & $\times$ & $\times$ & $\checkmark$ \\ \hline
\begin{tabular}[c]{@{}c@{}}Resistance to Man-in-\\ the Middle Attacks\end{tabular} & $\times$ & $\checkmark$ & $\checkmark$ & $\times$ & $\checkmark$ \\ \hline
Resistance to Replay Attacks & $\times$ & $\checkmark$ & $\checkmark$ & $\times$ & $\checkmark$ \\ \hline
\end{tabular}%
}
\end{table}

Table \ref{tab:Table5} shows a comparative summary of SACRIFICE with four state-of-the art papers \cite{WDU_TIFS'19,MEW_IoT'19,CHW_IoT'19,LLO_Systems'20} on the basis of key security features achieved by these schemes. It is evident that SACRIFICE outperforms all the other schemes considerably.

\section{Performance Analysis}

In this section, we evaluate the performance of SACRIFICE both theoretically and experimentally.

\subsection{Theoretical Analysis}

This section evaluates SACRIFICE in terms of its various overheads. It also analyzes the robustness of the scheme in terms of cracking probability.

\subsubsection{Overhead Analysis}

The computation, communication and storage overheads are measured in terms of execution time, number of transmitting and receiving bytes and number of bytes stored in the memory respectively. Table \ref{tab:Table1} summarizes the notations used. During analysis, we consider that the size of each element in $\mathbb{G}$ and $\mathbb{Z}_q^*$ of the elliptic curve is $128\ bytes$ and $20\ bytes$ respectively. We consider the size of timestamp variables (denoted as $|T|$) are $4\ bytes$. The overheads are calculated for a single round of mutual authentication, report generation etc. We also compare SACRIFICE with one competitor scheme \cite{WDU_TIFS'19}.

\begin{table}[!htb]
    \centering
    \caption{Notations used for Theoretical Analysis}
    \label{tab:Table1}
    \begin{tabular}{c|c}
        \hline
\textbf{Time taken for} & \textbf{Notation}\\
\hline
Scalar multiplication & $T_M$\\
Bilinear pairing & $T_{BP}$\\
Exponentiation & $T_E$\\
Hash operation & $T_H$\\
\hline
    \end{tabular}
\end{table}

\noindent \textbf{Computation Overhead:} Table \ref{tab:Table2} shows the computation overhead of SACRIFICE and the competing scheme. It is evident from the Table that the overhead for SACRIFICE is less than its competitor \cite{WDU_TIFS'19} for the first three phases. Additionally, SACRIFICE performs better than its competitor for the last two phases as well because of the absence of expensive operations like exponentiations, bilinear pairings etc. 

\begin{table}[!htb]
\centering
\caption{Comparative Analysis of Computation Overhead}
\label{tab:Table2}
\scalebox{0.77}{%
\begin{tabular}{|c|c|c|c|}
\hline
\textbf{Phase} & \textbf{Entity} & \textbf{SACRIFICE} & \textbf{Competing Scheme \cite{WDU_TIFS'19}} \\ \hline
\multirow{2}{*}{\begin{tabular}[c]{@{}c@{}}Device Registration\\ (Vehicle)\end{tabular}} & TA/SA & $T_H$ & $2T_M + 2T_E + T_H$ \\ \cline{2-4} 
 & Vehicle & $-$ & $T_M + 2T_{BP} + 2T_E + 2T_H$ \\ \hline
\multirow{2}{*}{\begin{tabular}[c]{@{}c@{}}Device\\ Registration (RSU)\end{tabular}} & TA/SA & $T_H$ & $2T_M + 2T_E + T_H$ \\ \cline{2-4} 
 & RSU & $-$ & $T_M + 2T_{BP} + 2T_E + 2T_H$ \\ \hline
\multirow{2}{*}{\begin{tabular}[c]{@{}c@{}}Mutual\\ Authentication\end{tabular}} & Vehicle & $3T_M + 3T_H$ & $3T_M + 2T_{BP} + 5T_E + 4T_H$ \\ \cline{2-4} 
 & RSU & $4T_M + 4T_H$ & $3T_M + 2T_{BP}+ 5T_E + 4T_H$ \\ \hline
\multirow{2}{*}{\begin{tabular}[c]{@{}c@{}}Report\\ Generation\end{tabular}} & Vehicle & $T_H$ & $T_M + 4T_E + 3T_H$ \\ \cline{2-4} 
 & RSU & $2T_M + 6T_H$ & $-$ \\ \hline
\multirow{2}{*}{\begin{tabular}[c]{@{}c@{}}Report\\ Processing\end{tabular}} & CS & $2T_M*n+T_H$ & $T_M+ (2n+2)T_{BP} + 3T_E + 3T_H$ \\ \cline{2-4} 
 & AA/RA & $2T_M + 3T_H$ & $4T_{BP} + 4T_E + 2T_H$ \\ \hline
\end{tabular}%
}
\end{table}

\noindent \textbf{Communication Overhead:} Table \ref{tab:Table3} provides communication and storage overhead analysis. From the Table, we observe that the communication overhead for SACRIFICE is significantly less than the overheads of the competitor. This is because SACRIFICE uses lightweight cryptographic tools like hash functions which reduces the size of the tokens exchanged during communication.

\noindent \textbf{Storage Overhead:} We observe from Table \ref{tab:Table3} that the storage overhead for SACRIFICE is also less than its competitor. Even though the RSU stores an additional $128\ bytes$ during mutual authentication, the overall storage overhead of SACRIFICE is still less than the work \cite{WDU_TIFS'19}.

\begin{table}[!htb]
\centering
\caption{Comparative Analysis of Communication and Storage Overheads}
\label{tab:Table3}
\scalebox{0.60}{%
\begin{tabular}{|c|c|c|c|c|c|c|c|}
\hline
\multirow{3}{*}{\textbf{Phases}} & \multirow{3}{*}{\textbf{Entity}} & \multicolumn{4}{c|}{\textbf{Communication Overhead (bytes)}} & \multicolumn{2}{c|}{\textbf{Storage Overhead (bytes)}} \\ \cline{3-8} 
 &  & \multicolumn{2}{c|}{\textbf{SACRIFICE}} & \multicolumn{2}{c|}{\textbf{Competing Scheme \cite{WDU_TIFS'19}}} & \multirow{2}{*}{\textbf{SACRIFICE}} & \multirow{2}{*}{\textbf{\begin{tabular}[c]{@{}c@{}}Competing\\ Scheme \cite{WDU_TIFS'19}\end{tabular}}} \\ \cline{3-6}
 &  & \multicolumn{1}{l|}{\textbf{Transmitted}} & \multicolumn{1}{l|}{\textbf{Received}} & \multicolumn{1}{l|}{\textbf{Transmitted}} & \multicolumn{1}{l|}{\textbf{Received}} &  &  \\ \hline
\multirow{2}{*}{\begin{tabular}[c]{@{}c@{}}Mutual\\ Authentication\end{tabular}} & Vehicle & 280 & 152 & 556 & 540 & 20 & 384 \\ \cline{2-8} 
 & RSU & 152 & 280 & 540 & 556 & 20 & 384 \\ \hline
\multirow{3}{*}{\begin{tabular}[c]{@{}c@{}}Report\\ Generation\end{tabular}} & Vehicle & 44 & $-$ & 992 & $-$ & 404 & 772 \\ \cline{2-8} 
 & RSU & 216 & 44 & $-$ & $-$ & 128 & $-$ \\ \cline{2-8} 
 & CS & $-$ & 216 & $-$ & 992 & $-$ & $-$ \\ \hline
\multirow{2}{*}{\begin{tabular}[c]{@{}c@{}}Report\\ Processing\end{tabular}} & CS & 216 & $-$ & 992 & $-$ & $n \cdot$ 216 & $n \cdot$ 772 \\ \cline{2-8} 
 & AA/RA & $-$ & 216 & $-$ & 992 & $-$ & $-$ \\ \hline
\end{tabular}%
}
\end{table}

Summarily, we observe from the entire overhead analysis that SACRIFICE is less intensive in terms of computation, communication and storage overheads than its competitor. This is because a less intensive mathematical approach is used in our proposed scheme. However, this approach maybe vulnerable to brute-force attacks, hence we explore the probability of breaking SACRIFICE by a brute-force attack in the following subsection.

\subsubsection{Cracking Probability}

Cracking probability is defined as the probability of cracking a token while it is being transmitted or stored by an adversary \cite{CDB_15}. Mutual Authentication is a crucial step in SACRIFICE. If an attacker breaks this step, it can gain control over the network and its confidential information. It can also send malicious data to other participants by compromising them. The following cases describe the probability of cracking the mutual authentication algorithm when an adversary launches a brute-force attack on the network.

\noindent \textbf{Case 1:} When an attacker $\mathcal{A}$ attempts to generate message $M_1=\{X_1,X_3,C_i,t_{U_i}\}$, by posing as an authentic vehicle in the network the following may happen:

\noindent The tokens $X_1$, $X_3$ can be easily calculated by $\mathcal{A}$ with any random number $r_i \in Z_q^*$  and some identity $ID_{U_i}$. However, for $C_i$ it does not have the value $P_{U_i}$ (since the vehicle is not registered). Therefore, $\mathcal{A}$ has to select a value for $C_i$.

\noindent The length of the token = $|C_i| = |Z_q^*| = n$ bits (Say)\\
Total possible combinations of the $C_i$ bits = $2^n$\\
Probability that the correct combination for this case is selected = $\frac{1}{2^n}$\\
Therefore, the Cracking Probability (\%) $= \frac{1}{2^n} \times 100$

\noindent For example, for $|Z_q^*|$ = 64 bits, the Cracking Probability (\%) = $5.4*10^{-18}$ is very low.

\noindent \textbf{Case 2:} When an attacker $\mathcal{A}$ attempts to generate message $M_2=\{Y_1,C_j,t_(R_j )\}$, by posing as an authentic RSU in the network, the following may occur:

\noindent The tokens $Y_1$ can be easily calculated by $\mathcal{A}$ with any random number $r_j \in Z_q^*$. However, it does not have the necessary tokens to calculate the value of $C_j$. Therefore, $A$ has to select a value for $C_j$ and similar to \textit{Case 1}, the Cracking Probability (\%) = $\frac{1}{2^n} \times 100$.

\noindent Thus, it is clear from the above discussion that due to the very low cracking probability, an attacker $\mathcal{A}$ cannot infiltrate the system by a brute-force attack within the short span of time that a particular vehicle stays in the range of an RSU. 

\subsection{Experimental Evaluation via Simulation}

Here, we implement a prototype of SACRIFICE and validate its performance with a state-of-the art competitor \cite{WDU_TIFS'19}.

\subsubsection{Simulation Environment}

\begin{figure}[!htb]
    \begin{minipage}[t]{.48\linewidth}
        \centering
        \includegraphics[width=\textwidth, height=1in]{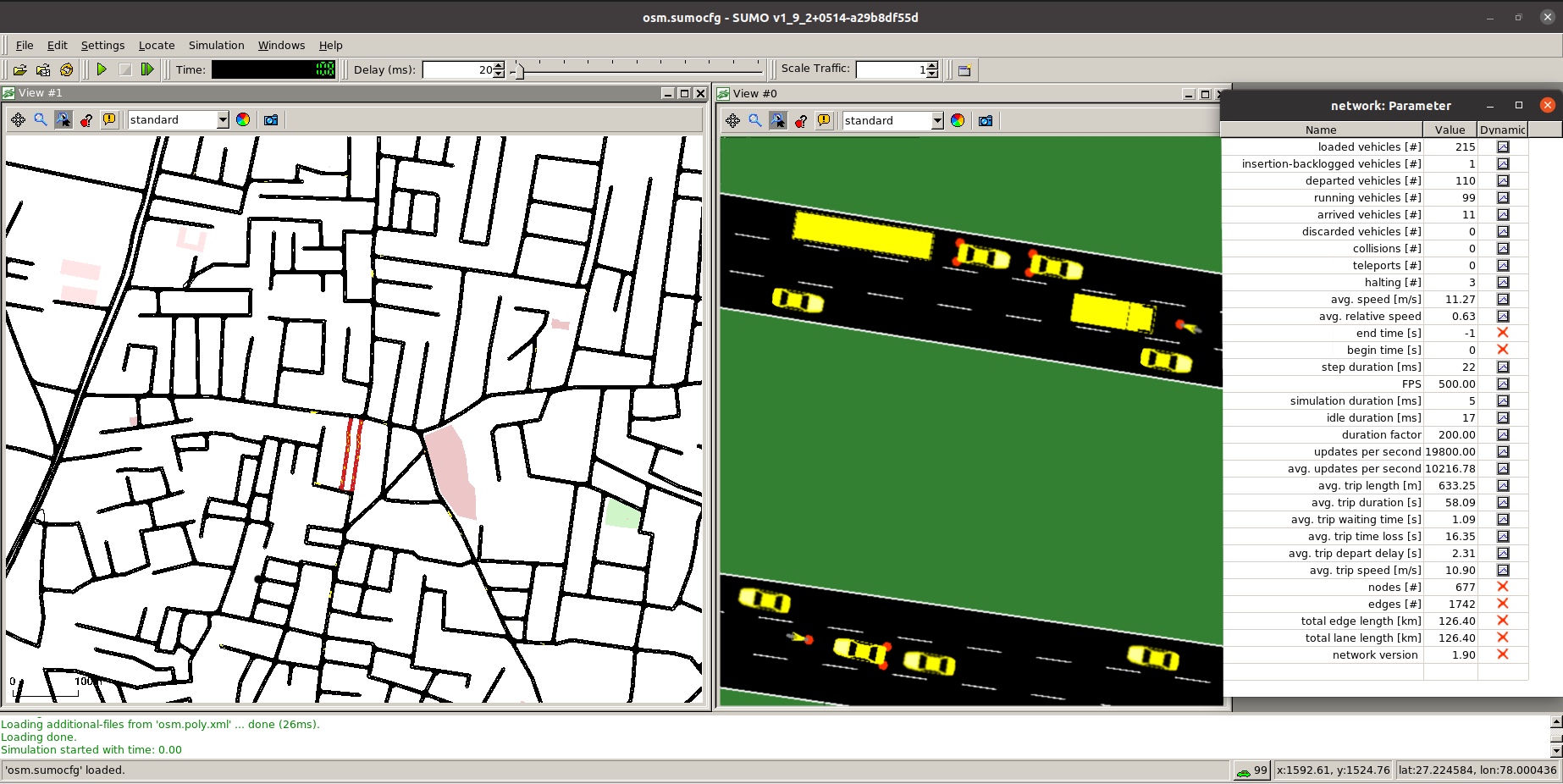}
        \subcaption{SUMO} \label{fig:Image5a}
    \end{minipage}
   \hspace{0.2cm}
    \begin{minipage}[t]{.48\linewidth}
        \centering
        \includegraphics[width=\textwidth, height=1in]{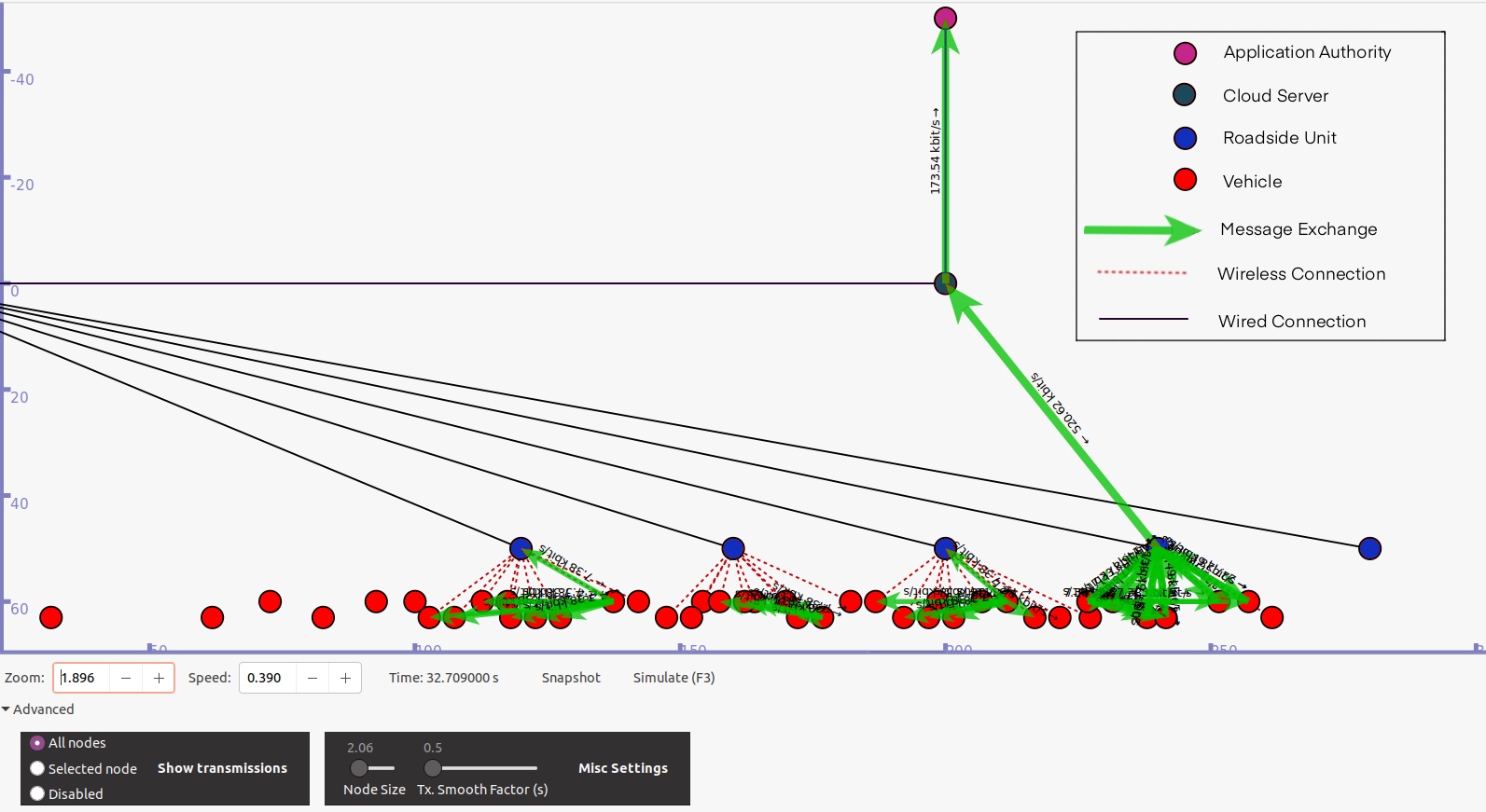}
        \subcaption{NS-3}\label{fig:Image5b}
    \end{minipage}
    \label{fig:Image5}
    \caption{\small \sl Simulation Snapshots}
\end{figure}

We implement SACRIFICE and its competitor with the help of two simulators: SUMO and NS3. For both the schemes, the pairing and group related operations are performed using the popular \textit{PBC Library} \cite{PBC} of NS-3, where we have used Type A pairings based on the elliptic curve, $y^2=x^3+x$. Table \ref{tab:Table4} summarizes the simulation parameters used in our setup. The simulation scenario in SUMO consists of a single street, 1000 m long, with 2 lanes shown in Fig. \ref{fig:Image5a}. The RSUs are deployed at an equal distance of 200 m from each other to provide maximum coverage. Fig. \ref{fig:Image5b} shows a snapshot of the simulation in NS3 taken at a particular time, $t=32.709$ seconds. It shows $42$ vehicles in the network (represented as red circles) which interact with the RSUs (represented as blue circles). It can be seen that some vehicles are interacting with the RSUs while a few have moved out of scope of all RSUs and others are still to enter the scope of any RSU. It is also observed that an RSU is sending a final road condition report to the CS which then forwards it to the Application Authority.

\begin{table}[!t]
    \centering
    \caption{Simulation Parameters}
    \label{tab:Table4}
    \begin{tabular}{c|c}
    \hline
\textbf{Parameters} & \textbf{Value}\\
        \hline
Area (SUMO) & $200 \times 1000$ $m^2$\\\hline
Duration (NS-3) & 5 minutes\\\hline
Range of Entities (NS-3) & 300 meters\\\hline
\multirow{2}{*}{Wireless Protocols (NS-3)} & 802.11p for message transmission\\
& 802.11b for beacon broadcasting\\
\hline
    \end{tabular}
\end{table}

\begin{figure*}[htb]
\begin{minipage}[b]{0.30\linewidth}
\centering
\includegraphics[width=\textwidth, height=1in]{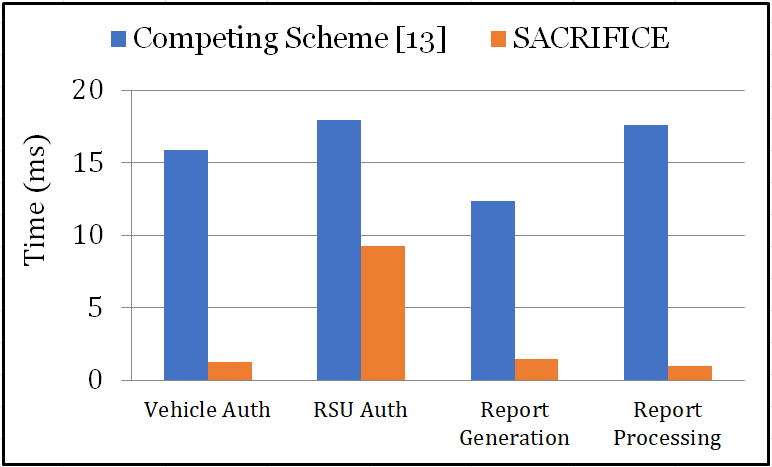}
\caption{\small \sl Performance Comparison of various sub-tasks}
\label{fig:Image8}
\end{minipage}
\hspace{0.5cm}
\begin{minipage}[b]{0.30\linewidth}
\centering
\includegraphics[width=\textwidth, height=1.1in]{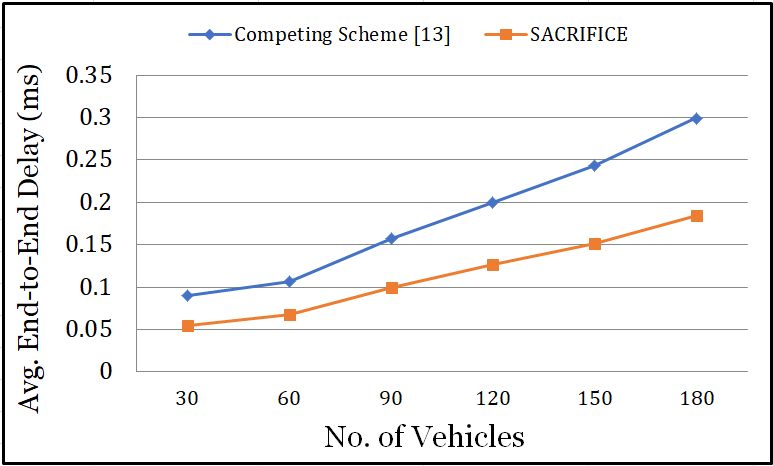}
\caption{\small \sl Average End-to-End Delay}
\label{fig:Image9}
\end{minipage}
\hspace{0.5cm}
\begin{minipage}[b]{0.30\linewidth}
\centering
\includegraphics[width=\textwidth, height=1.1in]{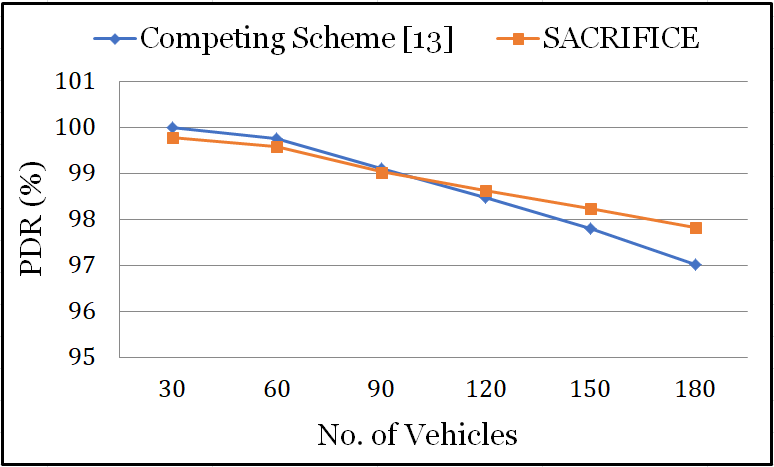}
\caption{\small \sl Packet Delivery Ratio} 
\label{fig:Image10}
\end{minipage}
\end{figure*}

\subsubsection{Simulation Metrics} We measure the performance of SACRIFICE primarily by evaluating the efficiency of it in terms of time taken both for each of the sub-tasks and entities involved in the scheme. Apart from this, we measure underlying network performance in implementing the scheme. The following two metrics are used in evaluating such network performance:

\noindent \textbf{End-to-end Delay:} Delay is defined as the average time taken since the moment a vehicle starts transmitting the packet until its successful delivery \cite{Sensors'17}. In our work, we consider the end-to-end delay as the average delay of all the packets transmitted in the network within the simulation duration.
$$Avg.\ Delay \: = \: \frac{sum \: of \: delay \: of \: all \: packets \: received}{total \: no. \: of \: packets \: received}$$
\noindent \textbf{Packet Delivery Ratio (PDR):} It is measured as the ratio of the total number of packets delivered to the destination to the total number of packets sent to the destination over a period of time \cite{VSD_iSES'18,KSD_21}. 
$$PDR \: = \: \frac{total \: no. \: of \: packets \: delivered}{total \: no. \: of \: packets \: sent}$$

\subsubsection{Results and Discussion} We conduct four sets of experiments to evaluate the performance of SACRIFICE and its competitor scheme \cite{WDU_TIFS'19}. In the simulation, vehicles start to send their own authentication messages after they receive the beacon broadcast from the nearest RSU. Such broadcasts happen every 0.3 milliseconds. An average result of 10 independent runs is taken while plotting the simulation graphs.

In the first set of experiment, we plot (Fig. \ref{fig:Image8}) the execution time of the various sub-tasks. We observe that the time taken by each sub-task in SACRIFICE is substantially less compared to its state-of-the-art competitor \cite{WDU_TIFS'19}. SACRIFICE roughly takes around $80.96\%$ on average less execution time for each of the sub-tasks compared to its competitor scheme.

% \begin{figure}[!htb]
% \centering
% \includegraphics[width=0.9\linewidth]{Graphs/Graph1.PNG}
% \caption{Performance Comparison of various sub-tasks}
% \label{fig:Image8}
% \end{figure}

In the second set of experiment, we observe (Table \ref{tab:Table6}) the time taken by each entity in SACRIFICE is less compared to that of its competitor. We also observe that the the time taken is of the order of a few nanoseconds and hence is feasible for time-critical VANET applications. As explained earlier, the reason that SACRIFICE performs better in both these set of experiments is because it uses less intensive mathematical operations as compared to its competitor.

\begin{table}[!ht]
\centering
\caption{Average Time taken by each Entity}
\label{tab:Table6}
\scalebox{0.80}{%
\begin{tabular}{|c|c|c|}
\hline
\textbf{Entity} & \textbf{SACRIFICE (ns)} & \textbf{Competing Scheme \cite{WDU_TIFS'19} (ns)} \\ \hline
Vehicle & 3.552 & 34.176 \\ \hline
RSU & 9.31 & 28.13 \\ \hline
CS & 0.58 & 9.74 \\ \hline
AA/RA & 0.566 & 5.45 \\ \hline
\end{tabular}%
}
\end{table}

In the third set of experiment, we plot (Fig. \ref{fig:Image9}) the average end-to-end delay with increasing number of vehicles in the network. From the figure, it is evident that the average delay increases when the number of vehicles increases in the network. The increase in the number of vehicles results in increased number of packet transmission resulting in increased congestion. This network congestion, in turn, increases the end-to-end delay of each packet. On an average, the delay for SACRIFICE shows 37.5\% better performance as compared to the work \cite{WDU_TIFS'19}. 

% \begin{figure}[!htb]
% \centering
% \includegraphics[width=0.9\linewidth]{Graphs/Graph2.PNG}
% \caption{Average End-to-End Delay}
% \label{fig:Image9}
% \end{figure}

In the fourth set of experiment, we plot (Fig. \ref{fig:Image10}) the packet delivery ratio (\%) with increasing number of vehicles in the network. We observe from the figure that for both the schemes there is a negative trend in the graph, i.e. PDR decreases with the increasing number of vehicles in the network, as expected. However, the ratio is stable and stays above 97\% for both the schemes. We also observe that as the number of vehicles approaches 90, the PDR for the proposed scheme starts improving compared to its competitor. Thus, when the number of vehicles in the network is large, SACRIFICE scales well or performs better.

% \begin{figure}[!htb]
% \centering
% \includegraphics[width=0.9\linewidth]{Graphs/Graph3.PNG}
% \caption{Packet Delivery Ratio}
% \label{fig:Image10}
% \end{figure}

\section{Conclusion}

We propose a secure road condition monitoring scheme SACRIFICE over fog-based VANET, which is low-overhead and scalable. In this scheme, whenever a vehicle encounters a bad road condition (e.g. accident) in the network, it sends a report to the closest RSU only after performing mutual authentication between vehicle and RSU. The RSU then generates the final report and sends it to the CS for further processing. Apart from maintaining the important security like mutual authentication, user anonymity, the scheme also ensures additional security features like non-repudiation, unlinkability and untraceability. The detailed analysis of the security features shows that our scheme is robust against both external and internal adversaries. Performance of the scheme is evaluated both through theoretical overhead analysis and simulation using SUMO and NS-3 platform to show its viability for practical implementation. The overhead analysis shows our scheme’s dominance over a state-of-the-art competitor. Simulation results also corroborate the theoretical analysis in terms of execution time while achieving better network performance in terms of end-to-end delay thereby establishing its applicability in time-sensitive VANET application. In future, the scheme may be extended by including vehicle-to-everything (V2X) communications and its associated security issues. Introducing the concept of scheduling to improve the overall latency further is another open research area.

\bibliographystyle{unsrt}
{\footnotesize
\bibliography{references}}

\end{document}